\documentclass[prb,amsmath,amssymb,twocolumn,superscriptaddress]{revtex4-2}
\usepackage{graphicx}
\usepackage{epsfig}
\usepackage{dcolumn}
\usepackage{bbold}
\usepackage{bm}
\usepackage{rotating}
\usepackage{multirow}
\usepackage{xspace}
\usepackage[table]{xcolor}
\usepackage[english]{babel}
\usepackage{soul} 
\setstcolor{red}

\newcommand{\ket}[1]{ | #1 \rangle }
\newcommand{\bra}[1]{ \langle #1 | }
\DeclareMathAlphabet\mathbfcal{OMS}{cmsy}{b}{n}

\def\w{\omega}
\def\bk{{\bf k}}
\def\bkq{{{\bf k}+ {\bf q}}}

\def\br{{\bf r}}
\def\bq{{\bf q}}

\def\d{\dagger}
\def\>{\rangle}
\def\<{\langle}
\def\D{\partial}

\def\k{\kappa}

\begin{document}
\title{Quantum theory of light-driven coherent lattice dynamics} 
\author{Fabio Caruso}
\affiliation{Institut f\"ur Theoretische Physik und Astrophysik, Christian-Albrechts-Universit\"at zu Kiel, Kiel, Germany}
\affiliation{Kiel Nano, Surface and Interface Science KiNSIS, Kiel, Germany}
\author{Marios Zacharias}
\affiliation{ Univ Rennes, INSA Rennes, CNRS, Institut FOTON – UMR 6082, F-35000 Rennes, France}
\begin{abstract}
The exposure to intense electromagnetic radiation can induce distortions and
symmetry breaking in the crystal structure of solids, providing a route for the
all-optical control of their properties. In this manuscript, we formulate a
unified theoretical approach to describe the coherent lattice dynamics in
presence of external driving fields, electron-phonon and phonon-phonon
interactions, and  quantum nuclear effects.   The main mechanisms for the
excitation of coherent phonons -- including infrared absorption, displacive
excitation, inelastic stimulated Raman scattering, and ionic Raman scattering
-- can be seamlessly accounted for.  We apply this formalism to a model
consisting of two coupled phonon modes, where we illustrate the influence of
quantum nuclei on structural distortions induced by ionic Raman scattering.
Besides validating the widely-employed classical models for the coherent
lattice dynamics, our work provides a versatile approach to methodically
explore the emergence of quantum nuclear effects in the structural response of
crystal lattice to strong fields. 
\end{abstract}
\maketitle
\section{Introduction}
The structural dynamics of solids triggered by intense light pulses
\cite{kampfrath_resonant_2013} can act as a precursor for driving materials
across phase transitions \cite{torre21} or into metastable states inaccessible
in equilibrium conditions \cite{nova_metastable_2019}.  Besides the fundamental
importance of this route for light-assisted structural control, numerous
domains of application where these concepts could be of technological relevance
can be easily envisioned.  Experimental realizations of this paradigm are
numerous and include the all-optical switching of electric polarization in
ferroelectrics
\cite{mankowsky_ultrafast_2017,nova_metastable_2019,li_terahertz_2019},
insulator to metal transitions \cite{rini_control_2007}, enhanced
superconductivity
\cite{mankowsky_nonlinear_2014,mitrano_possible_2016,Demler/2016}, or the
tuning of spin dynamics \cite{Kampfrath_2010} and magnetism
\cite{nova_effective_2017,maehrlein_dissecting_2018,juraschek_magnetic_2021,juraschek_phono-magnetic_2020,afanasiev_ultrafast_2021,sharma_making_2022}.
A review of the recent progress in exploring and exploiting light-induced
structural control and related phenomena in matter can be found in
Refs.~\cite{torre21,disa_engineering_2021}.

\begin{figure}[t]
\begin{center}
\includegraphics[width=0.48\textwidth]{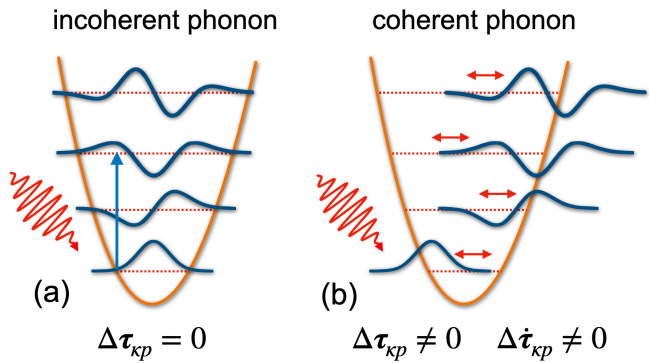}
\caption{\label{fig:coherent}
Schematic illustration of an incoherent (a) and coherent (b) phonon excitation.
Incoherent phonons are determined by the quantum number of the harmonic
oscillator $n_{\bq\nu} $ and maintain a zero average displacement of the
lattice ($\Delta{\boldsymbol \tau}_{\k p} =0 $).  Coherent phonons are
characterized by the displacement of the nuclear wave packets from thermal
equilibrium ($\Delta{\boldsymbol \tau}_{\k p} \neq 0 $) induced by external
perturbations which lead to a periodic motion of the nuclear wave packets
($\Delta\dot{\boldsymbol \tau}_{\k p} \neq 0 $).}
\end{center}
\end{figure}

The non-equilibrium dynamics of the crystalline lattice triggered by light
absorption consists simultaneously of {\it incoherent} and {\it coherent}
parts. The incoherent lattice dynamics is characterized by the increase in
phonon population of the phonon modes directly coupled to the field as, for
example, infrared (IR) active modes driven by absorption of a terahertz (THz)
field.  As schematically illustrated in Fig.~\ref{fig:coherent} (a), this
mechanism is equivalent to the excitation of the quantum harmonic oscillator
upon absorption of a photon.  The coherent dynamics, conversely, is linked to
the displacement of the nuclear wave packets from their equilibrium positions
[Fig.~\ref{fig:coherent} (b)], and it underpins the periodic motion of the
nuclei around their thermal equilibrium positions
\cite{zeiger_theory_1992,kuznetsov_theory_1994,garrett_coherent_1996,Merlin1997}.
This behaviour is absent in a purely incoherent state of the lattice (e.g., at
thermal equilibrium), where the nuclear vibrations depend on the phonon modes
population plus the associated zero-point amplitude.  Several coherent phonon
excitation mechanism are discussed in the literature, including THz light
pulses \cite{Merlin1997}, impulsive stimulated Raman scattering (ISRS)
\cite{DESILVESTRI1985146,garrett_coherent_1996,Merlin1997,Stevens_Coherent_2002},
ionic Raman scattering
\cite{Maradudin_IRS_1970,Maradudin_IRS_1971,Humphreys_IRS_1972,forst_nonlinear_2011},
sum-frequency Raman scattering \cite{Maehrlein_sumfreq_2017}, and sum-frequency
ionic Raman scattering \cite{Juraschek2018}.  

A theoretical description of the incoherent lattice dynamics can be addressed,
for instance, from the solution of the time-dependent Boltzmann equation (TDBE)
for the lattice \cite{caruso2021,Bernardi2021} by accounting explicitly for
coupling to external fields, as well as the scattering processes  that drive
the system towards thermalization (electron-electron, electron-phonon,
phonon-phonon).  In the TDBE formalism, the non-equilibrium lattice dynamics is
approximately described by the time-dependent changes of the phonon occupations
arising from external perturbations \cite{Caruso/Novko/review}.  Overall, the
TDBE suffices to describe the non-equilibrium variations of phonon occupations
[Fig.~\ref{fig:coherent} (a)] associated with the incoherent dynamics
\cite{seiler_accessing_2021,BrittNL22}, however, it is inherently unsuitable to
account for coherent structural dynamics. 

If the quantum nature of the nuclei is neglected, the coherent lattice dynamics
in presence of external driving fields can be formulated within the framework
of classical mechanics \cite{shen_theory_1965} and recast in the form of a
driven harmonic oscillator equation of motion (EOM), which admits a numerical
or analytical solution.  Classical models have seen wide application in the domain
of light-driven structural control
\cite{subedi_theory_2014,mankowsky_nonlinear_2014,
subedi_proposal_2015,Fechner2016,subedi_midinfrared-light-induced_2017,
juraschek_dynamical_2017,Juraschek2017,nova_effective_2017,Juraschek2018,
juraschek_orbital_2019,nova_metastable_2019,juraschek_parametric_2020,juraschek_cavity_2021},
and they have proven suitable and predictive in describing physical phenomena
in which the structural dynamics can be attributed to two or few coupled
phonons.  A limitation of driven harmonic oscillator models is that they rely on a classical
description of nuclear motion, which makes them unsuitable to assess the
influence of quantum nuclear effects on the coherent lattice dynamics.
Additionally, in their most common implementations, only lattice dynamics in
the unit cell is considered although this can be circumvented within a
supercell approach.  Quantum  effects can play a substantial role in the
nuclear dynamics at low temperatures and/or in compounds with light atoms
(e.g., H and Li) \cite{Rossi/21,Grandi/PRB/2021}; the emergence of long-range
order (e.g. charge-density waves, disorder, or domain formation) are inherently
linked to the structural dynamics on length scales beyond the unit cell. 

Ab-initio molecular dynamics (AIMD) circumvents some of the limitations
mentioned above.  It can be extended to account for coupling to external
driving fields fully ab initio \cite{Bowman2003}; it accounts for lattice
anharmonicity up to all orders via the explicit calculation of the potential
energy surface for each nuclear configuration; the influence of phonons with a
finite wavevector on the dynamics can be encoded in the simulations by
extending the size of simulation cells.  However,  AIMD remains inherently
classical and it does not enable a simple inclusion of quantum nuclear effects.
In earlier works, quantum nuclear effects have been accounted for within the
framework of path-integral molecular dynamics \cite{Marx1996}, or via the
introduction of suitable thermostats \cite{Ceriotti2009}. Additionally, AIMD
simulations in supercells remain prohibitively expensive for all but the
simplest systems. Overall, these considerations outline the limitations of
existing theoretical approaches in the description of the light-driven
structural control. 

In this manuscript, we formulate a quantum theory of the light-induced coherent
structural dynamics in solids.  Our approach is based on the solution of the
Heisenberg EOM for the nuclear coordinate operator, and  allows
for systematically including (i) the coupling to external fields radiation at
linear order in the field intensity; (ii) quantum nuclear effects; (iii)
lattice anharmonicities due to all phonons. We show that the classical models, widely
employed in the field of non-linear phononics, can be recovered within this
framework by neglecting quantum nuclear effects.  To corroborate these
considerations and explore the main features of the classical and quantum
dynamics, we solve numerically the lattice EOM for a
minimal model consisting of two phonons coupled by lattice anharmonicities and
driven by a THz field.

The manuscript is organized as follows.  In Sec.~\ref{sec:H} we introduce the
general Hamiltonian for a crystalline lattice.  In Sec.~\ref{sec:TDBE} we
discuss the TDBE and its application to the study of incoherent phonons.
Section~\ref{sec:EOM} discusses a general approach to derive the coherent
dynamics of solids based on the solution of the Heisenberg EOM for the nuclear
displacement operator.  Section~\ref{sec:F} discusses the different forms of
the coherent phonon driving forces.  Section~\ref{sec:IR} illustrates 
application to the coherent dynamics of a harmonic lattice interacting with a
THz field.  In Sec.~\ref{sec:PP} the lattice dynamics in the presence of
anharmonicities is discussed.  In Sec.~\ref{sec:model} we demonstrate the
present formalism using a two-phonon model. Summary and conclusions are
presented in Sec.~\ref{sec:conc}.  

\section{Ab-initio Lattice Hamiltonian}\label{sec:H}
The Hamiltonian for an anharmonic lattice interacting with an electromagnetic
field can be expressed as \cite{bruesch1982phonons}:
\begin{align}\label{eq:H} \hat{H} &= \hat{H}_{\rm ph} + \hat{H}_{\rm eph}  +
\hat{H}_{\rm pp}  +  \hat{H}_{\rm IR}, \end{align}
where $\hat{H}_{\rm ph}$,  $\hat{H}_{\rm eph}$,  $\hat{H}_{\rm pp}$, and
$\hat{H}_{\rm IR}$ represent the Hamiltonian of the lattice in the harmonic
approximation, the electron-phonon and phonon-phonon interaction Hamiltonians,
and the Hamiltonian accounting for the coupling with the THz field,
respectively.  In second-quantization, $\hat{H}_{\rm ph}$ can be expressed  as: 
\begin{align}\label{eq:Hph} \hat{H}_{\rm ph} &= \sum_{\bq\nu} \hbar\w_{\bq\nu}
(\hat{a}_{\bq\nu}^{\d} \hat{a}_{\bq\nu} +1/2)  \quad.  \end{align}
Here, $\w_{\bq\nu}$ is the frequency of a phonon with momentum $\bq$ and mode
index $\nu$.  The sum over $\bq$ extends over crystal momenta in the Brillouin
zone.  $\hat{a}_{\bq\nu}^{\d}$ and $\hat{a}_{\bq\nu}$ are bosonic creation and
annihilation operators, respectively, which obey the usual commutation
relations.  In the following, we find it convenient to introduce the shortened
notation:
\begin{align} \label{eq:Qdef}
\hat{Q}_{\bq\nu} &= \hat{a}_{\bq\nu} + \hat{a}^\d_{-\bq\nu} \quad,\\
\hat{P}_{\bq\nu} &= \hat{a}_{\bq\nu} - \hat{a}^\d_{-\bq\nu} \quad.
\end{align}
The expectation values ${Q}_{\bq\nu} = \langle \hat{Q}_{\bq\nu} \rangle $ and
${P}_{\bq\nu} = \langle \hat{P}_{\bq\nu} \rangle $ yield the displacement
amplitude and momentum for a coherent phonon with momentum ${\bq}$ and index
$\nu$.  The commutation relations for $\hat{Q}_{\bq\nu}$ and $\hat{P}_{\bq\nu}$
are listed in Appendix \ref{sec:commutators}. 

$ \hat{H}_{\rm pp}$ is the phonon-phonon interaction Hamiltonian which accounts
for anharmonicities of the lattice.  Up to fourth order in the expansion of the
potential energy surface with respect to the lattice displacements, $
\hat{H}_{\rm pp}$ is given by: 
\begin{align}\label{eq:Hpp} &\hat  H_{\rm pp}= \frac{1}{3!} \sum_{\bq\bq'\bq''}
\sum_{\nu\nu'\nu''} \Psi^{(3)}_{\substack{\nu\nu'\nu'' \\ \bq\bq'\bq''}}
\hat{Q}_{\bq\nu}\hat{Q}_{\bq'\nu'}\hat{Q}_{\bq''\nu''} \\ &+ \frac{1}{4!}
\sum_{\bq\bq'\bq''\bq'''} \nonumber \sum_{\nu\nu'\nu''\nu'''}
\Psi^{(4)}_{\substack{\nu\nu'\nu''\nu''' \\ \bq\bq'\bq''\bq'''}}
\hat{Q}_{\bq\nu}\hat{Q}_{\bq'\nu'}\hat{Q}_{\bq''\nu''} \hat{Q}_{\bq'''\nu'''},
\end{align}
$\Psi^{(3)}$ and $\Psi^{(4)}$ are the third- and fourth-order phonon-phonon
coupling matrix elements corresponding to the partial derivatives of the
potential energy surface with respect to the phonon modes.  $\Psi^{(3)}$ and
$\Psi^{(4)}$  can be evaluated entirely from first principles via finite
difference \cite{ShengBTE_2014,Togo2015,FourPhonon2022} or  via stochastic
approaches based on ab-initio molecular dynamics
\cite{Hellmann2011,Monacelli_2021}.  The electron-phonon coupling Hamiltonian
is given by \cite{GiustinoRMP}: 
\begin{align} \label{eq:Heph} \hat{H}_{\rm eph} &= N_p^{-\frac{1}{2}}
\sum_{mn\bk} \sum_{\bq\nu} g_{mn}^\nu(\bk,\bq) \hat c^\d_{m\bkq} \hat c_{n\bk}
\hat{Q}_{\bq\nu} \quad.  \end{align}
$\hat c_{n\bk}^\d$ and $\hat c_{n\bk}$ denote fermionic creation and
annihilation operators, respectively, $ g_{mn}^\nu(\bk,\bq)$ is the
electron-phonon coupling matrix element, and $N_p$ is the number of $q$-points.  

\begin{figure}[t]
\begin{center}
\includegraphics[width=0.38\textwidth]{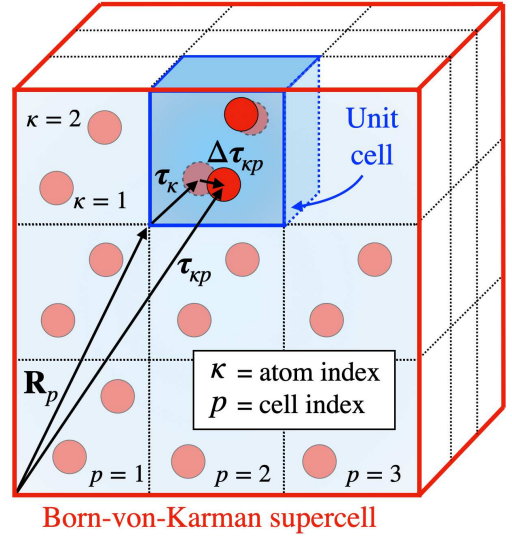}
\caption{\label{fig:BvK}
Schematic illustration of the Born-von-Karman supercell and of the notation
adopted to denote the nuclear displacements from equilibrium. }
\end{center}
\end{figure}

For electromagnetic fields with frequencies in the IR range,  the coupling to
the field can be expressed as: 
\begin{align} \label{eq:HIR} \hat{H}_{\rm IR} &= -e \sum_{\k p} {\bf E}(t)
\cdot {{\bf Z}^{\text{*}}_\k } \cdot \Delta \hat {\boldsymbol \tau}_{\k p}
\quad.  \end{align}
For sake of completeness, the derivation of Eq.~\eqref{eq:HIR} is reported in
Appendix~\ref{sec:HIR-deriv}.  ${{\bf Z}^{\text{*}}_\k }$  is the Born
effective-charge tensor \cite{GonzeLee1997,Baroni2001}, ${\bf E}(t)$ is a
time-dependent electric field, and $\Delta \hat{\boldsymbol \tau}_{\k p}$
denotes the displacement of the $\k$-th nucleus in the $p$-th unit cell from
its equilibrium configuration, which can be written as a linear combination of
normal modes \cite{GiustinoRMP}:
\begin{align}\label{eq:tau} \Delta \hat{\boldsymbol \tau}_{\k p} &=
\left(\frac{M_0}{N_p M_\k} \right)^{\frac{1}{2}} \sum_{\bq\nu} e^{i\bq \cdot
{\bf R}_p} {\bf e}^\k_{\bq\nu} l_{\bq\nu} \hat{Q}_{\bq\nu} \quad.  \end{align}
Here, $M_0$ is an arbitrary reference mass, $M_\k$ is the mass of the $\k$-th
nucleus, ${\bf e}^\k_{\bq\nu}$ are the phonon eigenvectors, and $l_{\bq\nu} =
\left({\hbar}/{2\w_{\bq\nu}M_0}\right)^{\frac{1}{2}}$ is the characteristic
length of a quantum harmonic oscillator with mass $M_0$ and frequency
$\w_{\bq\nu}$.  The position operator $\hat {\boldsymbol \tau}_{\k p}$ is
related to the displacement $ \Delta \hat{\boldsymbol \tau}_{\k p}$ by $\hat
{\boldsymbol \tau}_{\k p}  =  {\bf R}_{p} + \boldsymbol \tau_{\k} +  \Delta
\hat{\boldsymbol \tau}_{\k p}$, where ${\bf R}_{p}$ is a crystal-lattice
vector, and $ \boldsymbol \tau_{\k}$ is the equilibrium coordinate of the
$\k$-th nucleus in the unit cell, as illustrated schematically in
Fig.~\ref{fig:BvK}.

The operator $\Delta \hat {\boldsymbol \tau}_{\k p}$ is a key quantity for the
study of the lattice dynamics: its expectation value  $\Delta {\boldsymbol
\tau}_{\k p} (t) = \langle \Delta \hat{\boldsymbol \tau}_{\k p}(t) \rangle $
quantifies the displacement of a given nucleus from its equilibrium position
at time $t$, and  describes the coherent nuclear dynamics of the lattice.  In
the absence of a radiation filed, the eigenstates of a harmonic lattice are the
eigenstates $\ket{\chi_{ {n} }}$ of the Hamiltonian $\hat {H}_{\rm ph}$. They
satisfy $\bra {\chi_{n}} \hat{Q}_{\bq\nu} \ket{\chi_{{n}}} = 0 $, thus, leading
to vanishing displacements ${\boldsymbol \tau}_{\k p} (t) =0$. This is also the
case for incoherent phonons which do not contribute to the average displacement
of the nuclei.   It is clear from the definition
in Eq.~\eqref{eq:tau} that a prerequisite for having non-vanishing
displacements of the nuclear wave packets from equilibrium ($\Delta
{\boldsymbol \tau}_{\k p} \neq 0$) is the expectation value of the operator
$\hat Q_{\bq\nu}$ to be finite, i.e., $Q_{\bq\nu} = \langle \hat Q_{\bq\nu}
\rangle \neq 0$ \cite{kuznetsov_theory_1994}. Before proceeding to discuss how
this condition is realized, we briefly outline the TDBE and its application  
to the description of the incoherent lattice dynamics. 

\section{Incoherent phonons and the time-dependent Boltzmann equation}\label{sec:TDBE}
The TDBE is a well-established formalism to investigate the incoherent lattice
dynamics and determine the change of phonon population $n_{\bq\nu}$ for a
vibrating lattice subject to external perturbations \cite{Caruso/Novko/review}.
The influence of the electron-phonon coupling on the dynamics of phonons has
been subject of several ab-initio studies based on the TDBE approach
\cite{caruso2021,Bernardi2021,seiler_NL_2021,BrittNL22}.  In the following, we
briefly discuss the application of the TDBE to the lattice dynamics of polar
semiconductors interacting with a THz field.  In this case, all bands are
either filled or empty, and electron-phonon interactions are inconsequential
for the dynamics. The TDBE reads:  
\begin{align}\label{eq:BTEexact} \D_t n_{\bq\nu} = \Gamma^{\rm IR}_{\bq\nu}(t)
+ \Gamma^{\rm pp}_{\bq\nu}(t) \quad, \end{align}
with $\D_t = \D/\D t$.  Here, $\Gamma^{\rm pp}_{\bq\nu}$ is the scattering rate
due to phonon-phonon interactions, which can be obtained {\color {black} by
applying Fermi's golden rule to the Hamiltonian $\hat{H}_{\rm pp}$
\cite{Caruso/Novko/review}}.  Alternatively, it can be expressed in the
relaxation time approximation (RTA) as $\Gamma^{\rm pp}_{\bq\nu} = -(
n_{\bq\nu} - n_{\bq\nu}^{\rm eq}) / \tau_{\bq\nu}^{\rm pp}$ \cite{caruso2021},
where $n_{\bq\nu}^{\rm eq}$ denotes the equilibrium Bose-Einstein distribution,
and $\tau_{\bq\nu}^{\rm pp}$ is the relaxation time due to phonon-phonon
scattering \cite{ShengBTE_2014}.  $\Gamma^{\rm IR}_{\bq\nu}$ is the collision
integral due to IR absorption.  We show in Appendix~\ref{sec:appendix_gamma}
that it takes the following integral form: 
\begin{align}\label{eq:GammaIR} \Gamma^{\rm IR}_{\bq\nu}(t) =  e^2 E_0^2
&\frac{|{\bf F}_\nu \cdot {\boldsymbol \pi}|^2}{\hbar\omega_{\bq\nu}}
\delta_{\bq0} f(t)  \\ & \times \int_{-\infty}^t d\tau f(\tau) \cos[\w_{\bq\nu}
(\tau - t)] \nonumber\quad.  \end{align}
where ${\bf F}_\nu=  \sum_{\k} {
{{\bf Z}^{\text{*}}_\k } \cdot{\bf e}^\k_{0\nu} } \, M_{\k}^{-\frac{1}{2}}$ is
the IR cross section of the phonon $\nu$, also called mode effective charge.
Here, we express the electric field as ${\bf E}(t) = E_0 {\boldsymbol
\pi}f(t)$, where $E_0$ is the field intensity, $\boldsymbol \pi$ the
light-polarization unitary vector, and $f(t)$ an adimensional time-envelop
function. 

\begin{table}[b]
\setlength{\tabcolsep}{4.5pt}
\caption{
Computational parameters adopted in the investigation of the structural 
dynamics. 
}
\begin{tabular}{cccccc}
\hline\hline
                       $\w_{\rm IR} = \w$	& $\w_{\rm R}$ &  $E_0	$             & ${\bf Z}^*_\k$	& $M_\k	$ &  $| {\bf F}_\nu \cdot {\boldsymbol \pi}|$  \\ 
\hline                                                                    
    6~THz	        & 5~THz      & 100~kV~cm$^{-1}$     & $\pm6$           & 20~Da	&  0.05~$m_e^{-\frac{1}{2}}$    \\
\hline
\hline
\label{tab:param}
\end{tabular}
\end{table}

\begin{figure}[t]
\begin{center}
\includegraphics[width=0.42\textwidth]{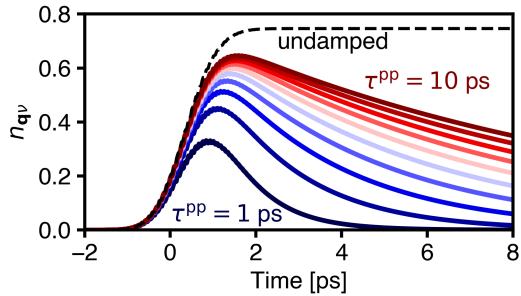}
\caption{\label{fig:TDBE_IR}
Change of phonon number for an IR-active phonon coupled to a THz field,
obtained from the solution of Eq.~\eqref{eq:BTEexact}. The computational
parameters are summarized in Table~\ref{tab:param}.  Phonon-phonon collisions
are treated in the relaxation time approximation by considering relaxation
times $\tau^{\rm pp}$ ranging between 1 and 10~ps.  The dynamics for the
undamped limit ($\tau^{\rm pp} \rightarrow \infty$) is marked by a dashed line.}
\end{center}
\end{figure}

To exemplify the (incoherent) lattice dynamics resulting from the solution of
Eq.~\eqref{eq:BTEexact}, we illustrate in Fig.~\ref{fig:TDBE_IR} the phonon
distribution function $n_{\bq\nu}$ for an IR-active mode with frequency
$\omega_{\rm IR} = 6$~THz coupled to a THz field with pulse profile $f(t) =
\sin(\omega t) e^{-(t/2\tau)^2}$ and resonant frequency $\omega = \omega_{\rm
IR}$.  The variation of the pulse profile $f(t)$ for durations $\tau = 0.1,
0.3,$ and $1.0$~ps is shown in Figs.~\ref{fig:IR}~(a-c).  To solve
Eq.~\eqref{eq:BTEexact} phonon-phonon scattering is accounted for in the RTA
with relaxation times $\tau^{\rm pp}_{\bq\nu}$ ranging between 1 and 10~ps and
for the undamped limit ($\tau^{\rm pp}_{\bq\nu}\rightarrow \infty$). We
considered a pulse duration $\tau = 2$~ps, and the field intensity is
parametrized by setting $E_0 = 100$~kV~cm$^{-1}$ and $| {\bf F}_\nu \cdot
{\boldsymbol \pi}|  = 0.05$~$m_e^{-\frac{1}{2}}$.  All parameters are
summarized in Table~\ref{tab:param}.  The trend illustrated in
Fig.~\ref{fig:TDBE_IR} exhibits an initial rise, which reflects the increased
phonon population due to IR absorption.  This mechanism is analogous to the
raise in quantum number of a driven quantum harmonic oscillator
[Fig.~\ref{fig:coherent}~(a)].  On longer timescales, the decrease of the
phonon number indicates the return to thermal equilibrium on timescales
dictated by phonon-phonon relaxation time $\tau^{\rm pp}$.  Different choices
for the field intensity $E_0$ or IR cross section ${\bf F}_\nu$ do not alter
this picture and lead to a simple rescaling of the $y$ axis in
Fig.~\ref{fig:TDBE_IR}. 

The dynamics of the phonon populations illustrated in Fig.~\ref{fig:TDBE_IR} is
a purely quantum effect of the lattice. In the classical picture, however, a
similar trend can be deduced by exploiting the correspondence between the
coherent vibrations and  the mean-square displacements of the ions
(Sec.~\ref{sec:QvsC}), and by employing phenomenological damping for the
phonon-phonon interaction
\cite{juraschek_orbital_2019,juraschek_parametric_2020}. 

\section{Equation of motion for the coherent lattice dynamics}\label{sec:EOM}

We proceed below to derive the general EOM for the coherent dynamics of the
lattice.  More precisely, we are interested in determining the time dependence
of the nuclear displacements $\Delta {\boldsymbol \tau}_{\k p} (t)$ for a
system described by the Hamiltonian $\hat{H} = \hat{H}_{\rm ph} + \hat{H}_{\rm
int}$, where $\hat{H}_{\rm int}$ denotes a generic perturbation.  In short, we
address this task by solving the Heisenberg EOM for the normal coordinate
operator $\hat{Q}_{\bq\nu}$ and deducing the nuclear displacements from
$\Delta {\boldsymbol \tau}_{\k p} (t)= \langle \Delta \hat{\boldsymbol
\tau}_{\k p} (t) \rangle$  via Eq.~\eqref{eq:tau}.  Here and below, the
Heisenberg picture is implied.  To illustrate these steps, we begin by
considering their application to an unperturbed harmonic lattice. 
 
\subsection{Coherent dynamics of the harmonic lattice. }\label{sec:eomHarm}
In an ideal harmonic lattice (i.e., for an Hamiltonian $\hat {H} = \hat
{H}_{\rm ph}$) the Heisenberg EOM for the normal coordinate $\hat{Q}_{\bq\nu}$, 
defined in Eq.~\eqref{eq:Qdef}, 
reads:
\begin{align}\label{eq:2der} i\hbar {\D_t \hat{Q}_{\bq\nu} } =  [
\hat{Q}_{\bq\nu}, \hat {H}_{\rm ph}  ] \quad, \end{align}
and similarly for $\hat{P}_{\bq\nu}$. Making use of the commutators in
Eqs.~\eqref{eq:cAHph} and \eqref{eq:cBHph} we promptly obtain: 
\begin{align} i\hbar {\D_t \hat{Q}_{\bq\nu} } &= \hbar\w_{\bq\nu}
\hat{P}_{\bq\nu} \quad, \\ i\hbar {\D_t \hat{P}_{\bq\nu} } &= \hbar\w_{\bq\nu}
\hat{Q}_{\bq\nu}  \quad.  \end{align}
By taking once more the time derivative, these expressions can be combined to
yield: 
\begin{align} {\D^2_t {Q}_{\bq\nu} } + \w_{\bq\nu}^2 {Q}_{\bq\nu} =0  \quad,
\end{align}
where we introduced the notation $ {Q}_{\bq\nu} =
\langle\hat{Q}_{\bq\nu}\rangle $, and $\langle \dots \rangle$ denotes the
expectation value taken with the initial state of the lattice.  This is the EOM
of a (classical) harmonic oscillator, which is solved by a linear combination
of complex exponentials with suitable initial conditions.  The time-dependent
nuclear displacements $\Delta {\boldsymbol \tau}_{\k p} (t)$ can be promptly
recovered through the transformation in Eq.~\eqref{eq:tau}.  However, because
$\langle \hat{Q}_{\bq\nu} \rangle =0$ the displacements vanish identically,
reflecting the absence of coherent atomic motion for an unperturbed harmonic
lattice. 

\subsection{Coherent dynamics in presence of interactions}
We next proceed to derive the general EOM for a harmonic lattice in presence of
an external perturbation $ \hat {H}_{\rm int}$, i.e., we consider the
Hamiltonian  $\hat {H} = \hat {H}_{\rm ph}+  \hat {H}_{\rm int}$.  By following
the procedure outlined in Sec.~\ref{sec:eomHarm} we arrive at the Heisenberg
EOM:
\begin{align} \label{eq:Atmp1} i\hbar {\D_t \hat{Q}_{\bq\nu} } &=
\hbar\w_{\bq\nu} \hat{P}_{\bq\nu}   \quad,\\ i\hbar {\D_t \hat{P}_{\bq\nu} } &=
\hbar\w_{\bq\nu} \hat{Q}_{\bq\nu} +  [ \hat{P}_{\bq\nu}, \hat {H}_{\rm int}  ]
\quad.  \label{eq:Btmp1} \end{align}
In Eq.~\eqref{eq:Atmp1}, we made use of $[ \hat{Q}_{\bq\nu}, \hat {H}_{\rm int}] =
0$,  which holds for the Hamiltonian introduced in Sec.~\ref{sec:H}
(see Appendix~\ref{sec:commutators}).
Equations~\eqref{eq:Atmp1} and \eqref{eq:Btmp1} can be recast into a
second-order differential equation for $ Q_{\bq\nu}$ by taking the time
derivative of Eq.~\eqref{eq:Atmp1} and its expectation value. This procedure 
leads to the general EOM for the coherent lattice dynamics in 
presence of external perturbations: 
\begin{align}\label{eq:eom_gen} {\D^2_t {Q}_{\bq\nu} } + \w_{\bq\nu}^2
Q_{\bq\nu} =  D_{\bq\nu}(t) \quad,  \end{align}
where we introduced the abbreviation: 
\begin{align}\label{eq:F} D_{\bq\nu}(t) = - \hbar^{-1}\w_{\bq\nu} \langle
[\hat{P}_{\bq\nu}, \hat {H}_{\rm int}  ]\rangle\quad.  \end{align}
Equation~\eqref{eq:eom_gen} assumes the familiar form of a driven harmonic oscillator
EOM. The quantity  $D_{\bq\nu}(t)$ acts as a time-dependent driving term  that
can induced finite coherent phonon amplitudes ${Q}_{\bq\nu}$  and a non-trivial
dynamics of the lattice.  
The precise form of the driving force $D_{\bq\nu}(t)$ and its time dependence 
depend on the nature of the interaction mechanism introduced by the
Hamiltonian $\hat {H}_{\rm int}$. The driving  term and its relation to the known mechanisms for 
the excitation of coherent phonons is discussed in more detail in Sec.~\ref{sec:F}. 

\subsection{Analytical solution for the coherent dynamics}\label{sec:an}
If lattice anharmonicities are neglected ($\hat H_{\rm pp}=0$), the EOM can be
solved analytically via the Green's function method. The coherent phonon
amplitudes can be recast in integral form as:
\begin{align}\label{eq:Q_exact} {Q}_{\bq\nu}(t)  = 
\int_{-\infty}^{t} G_{\bq\nu}(t,t')  D_{\bq\nu}(t') dt' \quad. \end{align}
Here, the Green's function is given by $G_{\bq\nu}(t,t') = 
{\sin[\w_{\bq\nu}(t-t')]}/\w_{\bq\nu}$, and  
the time-dependence of the nuclear coordinates 
$\Delta {\boldsymbol \tau}_{\k p}(t)$ can be recovered 
recovered through substitution of Eq.~\eqref{eq:Q_exact}
in Eq.~\eqref{eq:tau}. 
Equation~\eqref{eq:Q_exact} is the exact result 
for the coherent dynamics of an harmonic lattice. In this limit, the quantum and classical 
description coincide. 

\section{Driving mechanisms for the excitations of coherent phonons}\label{sec:F}
Below we discuss the form of the coherent phonon driving term in presence of 
electron-phonon coupling, external driving fields, and lattice anharmonicities. 
 This procedure enables to derive the known mechanisms for the excitation of 
coherent phonon within a unified theoretical framework. 

\subsection{Infrared absorption}\label{sec:DIR}
We begin by considering the IR driving mechanism, i.e., the direct excitation
of IR-active modes through the interaction with an electro-magnetic field with
frequency in the THz range.  We evaluate Eq.~\eqref{eq:F} using the Hamiltonian
$\hat{H}_{\rm IR}$ from Eq.~\eqref{eq:HIR} and making use of the  commutator in
Eq.~\eqref{eq:commIR}. One promptly obtains: 
\begin{align}\label{eq:DIR} D_{\bq\nu}^{\rm IR}(t) =   \delta_{\bq0} {\bf
C}_{\nu} \cdot {\bf E}(t) \quad.  \end{align}
Here, we defined ${\bf C}_\nu =   e N_p^{\frac{1}{2}} (2
\w_{0\nu}/\hbar)^\frac{1}{2} {\bf F}_\nu$. 
The time dependence of the coherent phonon driving force $D_{\bq\nu}^{\rm IR}$
is directly inherited from the electric field $ {\bf E}(t)$.  The quantity
$D_{\bq\nu}^{\rm IR}$  differs from zero only for zone-center  ($\Gamma$ point)
IR-active phonons. This follows directly from the dependence on the IR cross
section ${\bf F}_\nu$ and the negligible momentum of THz photons as compared to
the characteristic phonon momenta accessible in the Brillouin zone.
Correspondingly, only IR-active phonons at $\Gamma$ may exhibit coherent motion
in absence of lattice anharmonicities. 

\subsection{Electron-phonon coupling and the displacive excitation of coherent phonons}\label{sec:Deph}

We proceed to consider the driving mechanism arising from the electron-phonon
interaction, i.e., we consider the Hamiltonian  $\hat{H}_{\rm eph}$.  This
mechanism coincides with the so-called displacive excitation of coherent
phonons (DECP)
\cite{zeiger_theory_1992,kuznetsov_theory_1994,Sanders/PRB/2009,Lakehal/PRB/2019}.
Combining Eqs.~\eqref{eq:Heph}, \eqref{eq:F}, and \eqref{eq:commeph}, we arrive
at: 
\begin{align}\label{eq:Deph0} D_{\bq\nu}^{\rm eph} =  -\frac{ \omega_{\bq\nu}}
{\hbar N_p^{\frac{1}{2}}} \sum_{mn\bk}  g_{mn}^\nu(\bk,-\bq) \hat
c^\d_{m\bk-\bq} \hat c_{n\bk}\quad.  \end{align}
To eliminate the dependence on the fermionic operators $\hat c^\d_{m\bk-\bq}$
and $ \hat c_{n\bk}$, we take the expectation values using an electronic state
$\ket {\Psi}$.  Neglecting non-diagonal terms of the density matrix, i.e.,
$\bra{\Psi}  c^\d_{m\bk-\bq} \hat c_{n\bk} \ket {\Psi} \simeq f_{n\bk}(t)
\delta_{nm} \delta_{\bq,0}$  , one arrives at: 
\begin{align}\label{eq:Deph} D_{\bq\nu}^{\rm eph}(t) = -\frac{ \omega_{\bq\nu}}
{\hbar N_p^{\frac{1}{2}}} \sum_{n\bk}  g_{nn}^\nu(\bk,\bq)  [ f_{n\bk}(t) -
f_{n\bk}(0) ]  \delta_{\bq,0} \end{align}
where $ f_{n\bk}(t)$ denotes the electronic distribution function at time $t$.
The equilibrium ($t=0$) value $f_{n\bk}(0)$ has been subtracted because the
lattice is assumed to be initially at equilibrium.  Equation~\eqref{eq:Deph}
can be easily generalized to account for non-diagonal elements  of the density
matrix  \cite{kuznetsov_theory_1994}.  A straightforward symmetry analysis of
the electron-phonon coupling matrix elements $g_{nn}^\nu(\bk,\bq=0)$ reveals
that the driving force $D_{\bq\nu}^{\rm eph}(t)$ differs from zero only for
modes of $A_{1g}$ symmetry at the $\Gamma$ point.  However, coupling to lower
symmetry modes or to phonons with finite wavevectors ($\bq\neq0$) can arise if
the density matrix exhibits non-vanishing non-diagonal elements.  In short,
$D_{\bq\nu}^{\rm eph}(t)$ is a purely electronic driving mechanism that results
from the change of the electronic occupation function  $ f_{n\bk}(t) \neq
f_{n\bk}(0)$.  An example of this case the excitation of electrons via a pump
pulse. More generally, the DECP mechanism can persist even in absence of an
external driving field. In particular,  any physical process leading to a
sufficiently rapid change (i.e., faster than the characteristic decoherence
time) of the distribution function $f_{n\bk}$ should also lead to the
excitation of coherent phonons. 

\subsection{Inelastic stimulated Raman scattering}\label{sec:DISRS}

The direct excitation of coherent Raman-active phonons can arise via the
inelastic stimulated Raman scattering (ISRS) mechanism.  This process is the
only mechanism for the direct excitation of coherent phonons in crystals where
neither IR-active modes nor totally symmetric ($A_{1g}$) phonons are present
(e.g., diamond or graphene), and therefore the IR and displacive mechanisms
discussed in Secs.~\ref{sec:DIR} and \ref{sec:Deph}, respectively, are symmetry
forbidden.  The derivation of the driving force for ISRS is somewhat involved
and it requires the applications of second-order time-dependent perturbation
theory to the electron-phonon coupling driving force $D_{\bq\nu}^{\rm eph}$
[Eq.~\eqref{eq:Deph0}] to account for the effects of a off-resonance electric
field \cite{garrett_coherent_1996,Stevens_Coherent_2002}.  The resulting
driving force $D_{\bq\nu}^{\rm ISRS}$  depends quadratically on the field, and
it can be expressed within the Placzek approximation as
\cite{dhar_timeresolved_1994}: 
\begin{align}\label{eq:DISRS} D_{\bq\nu}^{\rm ISRS}(t) = \Omega
\sum_{\alpha\beta}  \frac{\partial \chi_{\alpha\beta} }{\partial Q_{\bq\nu}}
E_\alpha(t)  E_\beta(t) \end{align} where $\alpha$ and $\beta$ run over the
Cartesian components and $\chi_{\alpha\beta}$ is the linear susceptibility.
Earlier works suggested that the displacive mechanism and the corresponding
driving force  $D_{\bq\nu}^{\rm eph}$ can be regarded as a special case of the
ISRS mechanism \cite{garrett_coherent_1996}. 

\subsection{Phonon-phonon scattering and ionic Raman scattering}
Lattice anharmonicities yet provide an additional route for the coherent
excitations that do not couple directly to light.   An explicit expression for
the coherent phonon driving force $D_{\bq\nu}^{\rm 3pp}$ due to third-order
anharmonicities is derived by combining Eqs.~\eqref{eq:Hpp}, \eqref{eq:F}, and
\eqref{eq:commPP}, yielding:
\begin{align}\label{eq:Dpp3} D_{\bq\nu}^{\rm 3pp} (t) =
- \frac{\omega_{\bq\nu}}{\hbar } \sum_{\bq'\bq''} \sum_{\nu'\nu''}
  \Psi^{(3)}_{\substack{\nu\nu'\nu'' \\ -\bq\bq'\bq''}} \langle \hat Q_{\bq'
\nu' } \hat Q_{\bq'' \nu'' }\rangle \end{align}
where $\langle\cdots \rangle$ denote the expectation value with respect to the
initial state of the lattice, and its evaluation requires an approximate
treatment, as discussed in Secs.~\ref{sec:ppclass} and \ref{sec:ppquantum}.
Similarly, by considering fourth-order lattice anharmonicities (four-phonon
scattering processes), we arrive at:
\begin{align}\label{eq:Dpp4} D_{\bq\nu}^{\rm 4pp} (t) =
- \frac{\omega_{\bq\nu}}{3\hbar } \sum_{\substack{ \bq'\bq''\bq''' \\
  \nu'\nu''\nu'''}} \Psi^{(4)}_{\substack{\nu\nu'\nu''\nu''' \\
-\bq\bq'\bq''\bq'''}} \langle \hat Q_{\bq' \nu' } \hat Q_{\bq'' \nu''} \hat
Q_{\bq''' \nu'''}\rangle.  \end{align}
Lattice anharmonicities do not alter the coherent dynamics of the lattice at
equilibrium.  In presence of additional driving sources for the excitation of
coherent phonons, as, e.g., the mechanism discussed in
Secs.~\ref{sec:DIR}-\ref{sec:DISRS}, phonon-phonon scattering can lead to the
indirect excitations of coherent phonons that are coupled by lattice
anharmonicities.  The mechanism of ionic Raman scattering (IRS), for example,
is realized when a coherent IR phonon, photo-excited by IR absorption, drives
the excitation of a coherent Raman mode via phonon-phonon scattering.  IRS can
be accounted from first principles by simultaneously accounting for the IR and
the phonon-phonon scattering driving mechanisms $D_{\bq\nu}^{\rm IR}$ and
$D_{\bq\nu}^{\rm 3pp}$, respectively, in the EOM of the lattice.

\begin{figure*}[t] \begin{center}
\includegraphics[width=0.98\textwidth]{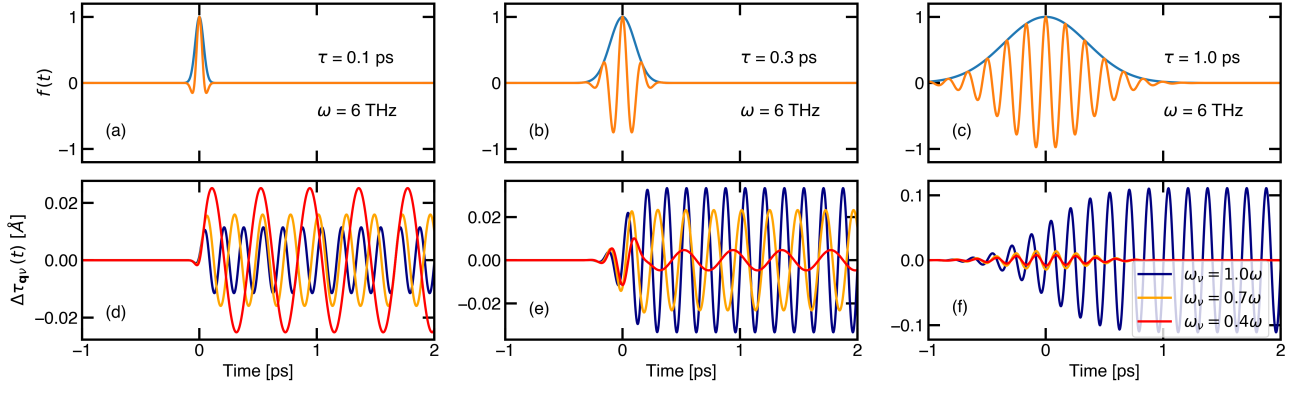} \caption{\label{fig:IR}
(a-c) Pulse profile $f(t) = \sin(\omega t) e^{-(2t/\tau)^2}$ (orange) for a
frequency $\omega=$6~THz and a Gaussian envelop function (blue) with width
$\tau=0.1$, 1, and 2 ps, respectively.  (d-f) Time-dependence of the function
$I_\nu (t) = \omega_\nu \int_{-\infty}^{t} {\sin[\w_{\nu}(t-t')]} f (t') dt' $
for phonon frequencies $\omega_\nu = 1.0\, \omega$, $ 0.7 \, \omega$,  and $
0.4 \,\omega$.  } \end{center} \end{figure*}

\section{Harmonic lattice in a THz field}\label{sec:IR}
In the following, we discuss the coherent dynamics of a harmonic crystal.  For
sake of conciseness, here and in the following we focus on the IR driving
mechanism ($D_{\bq\nu}^{\rm IR}$), although similar considerations can also be
extended, with due adjustments, to the displacive and Raman driving terms
($D_{\bq\nu}^{\rm eph}$ and  $D_{\bq\nu}^{\rm ISRS}$).  In this case, the EOM
for the lattice is promptly derived by combining the driving force
$D_{\bq\nu}^{\rm IR}(t)$ from Eq.~\eqref{eq:DIR}, with the generalized EOM from
Eq.~\eqref{eq:eom_gen}:
\begin{align}\label{eq:EOMIR} \frac{\D^2 {Q}_{\bq\nu}(t) }{\D t^2} +
\w_{\bq\nu}^2  Q_{\bq\nu}(t) = D_{\bq\nu}^{\rm IR} (t) \end{align}
Equation~\eqref{eq:EOMIR} can be solved numerically by ordinary
finite-difference algorithms for second-order differential equations (e.g.,
Heun or Runge Kutta).  The exact analytical solution of Eq.~\eqref{eq:EOMIR},
however, can also be obtained by direct analytical solution, as outlined in 
Sec.~\ref{sec:an}, yielding:  
\begin{align}\label{eq:tau-int-ir} \Delta {\boldsymbol \tau}_{\k p}(t) = 2 e
E_0 M_\k^{-\frac{1}{2}} & \sum_{\nu} {\bf e}^\k_{0\nu} ( \boldsymbol \pi \cdot
{\bf F}_{\nu} ) I_\nu(t)\omega_{0\nu}^{-2} \quad.  \end{align}
where $I_\nu (t)$ is an adimensional function of time  defined as:
\begin{align} I_\nu (t)  = \omega_{0\nu} \int_{-\infty}^{t}
{\sin[\w_{0\nu}(t-t')]} f (t') dt' \nonumber \quad.  \end{align}
and it encodes the time dependence of the nuclear displacements along the
$\nu$-th phonon mode.  Equation \eqref{eq:tau-int-ir} is the exact solution for
the coherent dynamics of a harmonic lattice in an IR field and it generalizes
the result obtained in Ref.~\cite{Merlin1997} for a single phonon mode. 

To illustrate the coherent dynamics triggered by a THz pulse, we report in
Figs.~\ref{fig:IR}~(d-f) the atomic displacement $\Delta {\boldsymbol \tau}_{\k
p}$ evaluated from the numerical integration of Eq.~\eqref{eq:tau-int-ir} for
the parameters listed in Table~\ref{tab:param}.  The field profile function
$f(t)$ is illustrated in Figs.~\ref{fig:IR}~(a-c) for a pulse duration of $\tau
= 0.1$, 1, and 2~ps and frequency $\omega=6$~THz.  In addition to the resonance
condition ($\omega=\omega_{0\nu}$), we further report numerical results for the
off-resonance case ($\omega_{0\nu} = 0.7\omega$ and $\omega_{0\nu} =
0.4\omega$). 

For impulsive perturbations  [Figs.~\ref{fig:IR}~(d)] -- namely, for pulse
durations shorter than the phonon period  ($\tau < 2\pi / \omega $)
[Figs.~\ref{fig:IR}~(a)]-- coherent nuclear motion (i.e., $\Delta {\boldsymbol
\tau}_{\k p}\neq0$) arises for both resonant ($\omega=\omega_{0\nu}$) and
non-resonant modes ($\omega\neq\omega_{0\nu}$).  In other words, the resonance
condition is not strictly enforced and also quasi-resonant modes can be driven
by a THz field.  For pulse durations comparable to or longer than the phonon
period ($\tau > 2\pi / \omega$) only resonant IR-active modes are excited
[Figs.~\ref{fig:IR}~(e-f)], whereas the dynamics of non-resonant modes is
strongly suppressed and it leads to a strict enforcement of the resonance
condition for long pulses. 

For times longer than the pulse duration ($t > 2 \tau$)  it is a good
approximation to substitute  $\int_{-\infty}^t$ with $\int^\infty_{-\infty}$
into Eq.~\eqref{eq:tau-int-ir}, and a fully analytical solution can be found
for the resonant case ($\omega= \omega_{0\nu}$): 
\begin{align}\label{eq:tau-int-ir2} \Delta {\boldsymbol \tau}_{\k p}(t) &=
\alpha \cos(\omega t) e E_0 M_\k^{-\frac{1}{2}} \sum_\nu {\bf e}^\k_{0\nu} (
\boldsymbol \pi \cdot {\bf F}_\nu ) \end{align}
where $\alpha = {\tau \sqrt{\pi}} (1- e^{-\frac{\omega^2 \tau^2}{4}})$, and the
sum extends over all resonant IR-active modes.  This expression provide a
simple tool to estimate the maximum displacements induced by a THz field in a
harmonic lattice.  In particular, in presence of a single resonant mode, the
maximum displacement of the nuclei induced by the field simply reduces to $
\Delta {\boldsymbol \tau}_{\k p}^{\rm max} = c_0 {\bf e}^\k_{0\nu}   $ with
$c_0= \alpha e E_0 M_\k^{-\frac{1}{2}}  ( \boldsymbol \pi \cdot {\bf F}_\nu )
$.

\section{Coherent dynamics of the anharmonic lattice}\label{sec:PP}
We thus proceed to investigate the effect of anharmonicities on coherent
lattice dynamics in presence of a THz driving field.  Replacing
Eqs.~\eqref{eq:DIR}, \eqref{eq:Dpp3}, and \eqref{eq:Dpp4} into
Eq.~\eqref{eq:eom_gen}, we obtain the general EOM for an THz-driven anharmonic
lattice: 
\begin{align}\label{eq:EOMpp} \D_t^2 & Q_{\bq \nu} + \omega_{\bq\nu}^2 Q_{\bq
\nu} = \delta_{\bq0} {\bf C}_\nu \cdot{\bf E}(t) \\&
- \frac{\omega_{\bq\nu}}{\hbar } \sum_{\bq'\bq''} \sum_{\nu'\nu''}
  \Psi^{(3)}_{\substack{\nu\nu'\nu'' \\ -\bq\bq'\bq''}} \langle \hat Q_{\bq'
\nu' } \hat Q_{\bq'' \nu'' }\rangle \nonumber \\&
- \frac{\omega_{\bq\nu}}{3\hbar } \sum_{\bq'\bq''\bq'''} \sum_{\nu'\nu''\nu'''}
  \Psi^{(4)}_{\substack{\nu\nu'\nu''\nu''' \\ -\bq\bq'\bq''\bq'''}} \langle
\hat Q_{\bq' \nu' } \hat Q_{\bq'' \nu''} \hat Q_{\bq''' \nu'''}\rangle
\nonumber \end{align}
This expression defines a coupled set of $N_p \times N_{\rm ph}$ differential
equations, with  $N_{\rm ph}$ being the number of phonons branches,
respectively.  While analytical solution is not possible, approximations can be
introduced to solve it numerically.  This can accomplished by: (i) neglecting
quantum effects (Sec.~\ref{sec:ppclass});  (ii) retaining only the lowest order
in the interaction (Sec.~\ref{sec:ppquantum}). 

\subsection{Classical approximation and  non-linear phononics models}\label{sec:ppclass}
A classical approximation to Eq.~\eqref{eq:EOMpp} can be deduced by neglecting
quantum correlations and thereby writing $ \langle \hat Q_{\bq' \nu' } \hat
Q_{\bq'' \nu'' } \rangle \simeq \langle \hat Q_{\bq' \nu' }\rangle \langle \hat
Q_{\bq'' \nu'' } \rangle =  Q_{\bq' \nu' }  Q_{\bq'' \nu'' } $ and similarly
for fourth-order anharmonicities. This approximation is equivalent to treat
nuclei as classical particles, and it is the most widely employed in the
description of coherent lattice dynamics.  Correspondingly, one arrives at the
classical EOM for the dynamics of an anharmonic lattice in a THz field: 
\begin{align}\label{eq:nlp} \D_t^2 &Q_{\bq\nu} + \omega_{\bq\nu}^2 Q_{\bq\nu} =
\delta_{\bq0} {\bf C}_\nu \cdot{\bf E}(t) \\ &-\frac{\omega_{\bq\nu}}{\hbar  }
\sum_{\nu'\nu''} \sum_{{\bq'\bq''}}
\Psi^{(3)}_{\substack{{\nu\nu'\nu''}\\{\bq\bq'\bq''}}}  Q_{\bq'\nu' }  Q_{
\bq''\nu'' } \nonumber \\ &- \frac{\omega_{\bq\nu}}{3\hbar }
\sum_{\bq'\bq''\bq'''} \sum_{\nu'\nu''\nu'''}
\Psi^{(4)}_{\substack{\nu\nu'\nu''\nu''' \\ \bq\bq'\bq''\bq'''}} Q_{\bq' \nu' }
Q_{\bq'' \nu''} Q_{\bq''' \nu'''} \nonumber \end{align}
Simplified versions of Eq.~\eqref{eq:nlp} are widely employed in non-linear
phononics to model the coherent structural dynamics driven by the absorption of
THz fields
\cite{subedi_theory_2014,juraschek_dynamical_2017,mankowsky_ultrafast_2017,juraschek_parametric_2020}.
Non-linear phononics models typically rely on the following approximations:
anharmonic coupling is restricted to two or few phonons; quantum nuclear
effects are neglected; only zone-center ($\Gamma$ point) phonons are
considered.  Some of these limitations can be easily lifted by noticing that
Eq.~\eqref{eq:nlp} is nothing else than the normal-coordinate representation of
the ordinary classical EOM of the lattice: 
\begin{align}\label{eq:md} M_\k  \Delta {\ddot  {\boldsymbol \tau}}_{\k p} =
\mathbfcal{F}_\k(t) + \frac{\D}{\D \Delta { {\boldsymbol \tau}}_{\k p}} U (\{
\Delta { {\boldsymbol \tau}}_{\k p} \}) \end{align}
where $ \mathbfcal{F}_\k(t) = e {\bf E} (t)  \cdot {\bf Z}_\k^{*}$ is the force
acting on $\k$-th nucleus due to the field ${\bf E}(t)$, and $U (\{ \Delta {
{\boldsymbol \tau}}_{\k p} \})$ is the potential energy surface expanded up to
fourth order in the nuclear displacements.  As long as lattice anharmonicities
are included at the same order, Eqs.~\eqref{eq:nlp} and \eqref{eq:md} are
completely equivalent formulations of the lattice dynamics in normal and
Cartesian coordinates, respectively.  Additionally, one can reversibly change
representation from normal to Cartesian coordinates (and {\it viceversa}) via
Eq.~\eqref{eq:tau} and the inverse transformation: 
\begin{align} \label{eq:NC2} Q_{\bq\nu} =  \sum_{\k p}
\left(\frac{M_\k}{M_0N_p}\right)^{\frac{1}{2}} e^{-i {\bq}\cdot {\bf R}_p}
l_{\bq\nu}^{-1} [{\bf e}^\k_{\bq\nu} ]^*\cdot  \Delta { {\boldsymbol \tau}}_{\k
p}\quad.  \end{align}
Equation~\eqref{eq:md} can be efficiently solved via AIMD by directly computing
the gradients of the potential-energy surface $U$ from Kohn-Sham
density-functional theory for a given displaced nuclear configuration.

\subsection{Quantum nuclear effects} \label{sec:ppquantum}

\begin{figure*}[t]
\begin{center}
\includegraphics[width=0.98\textwidth]{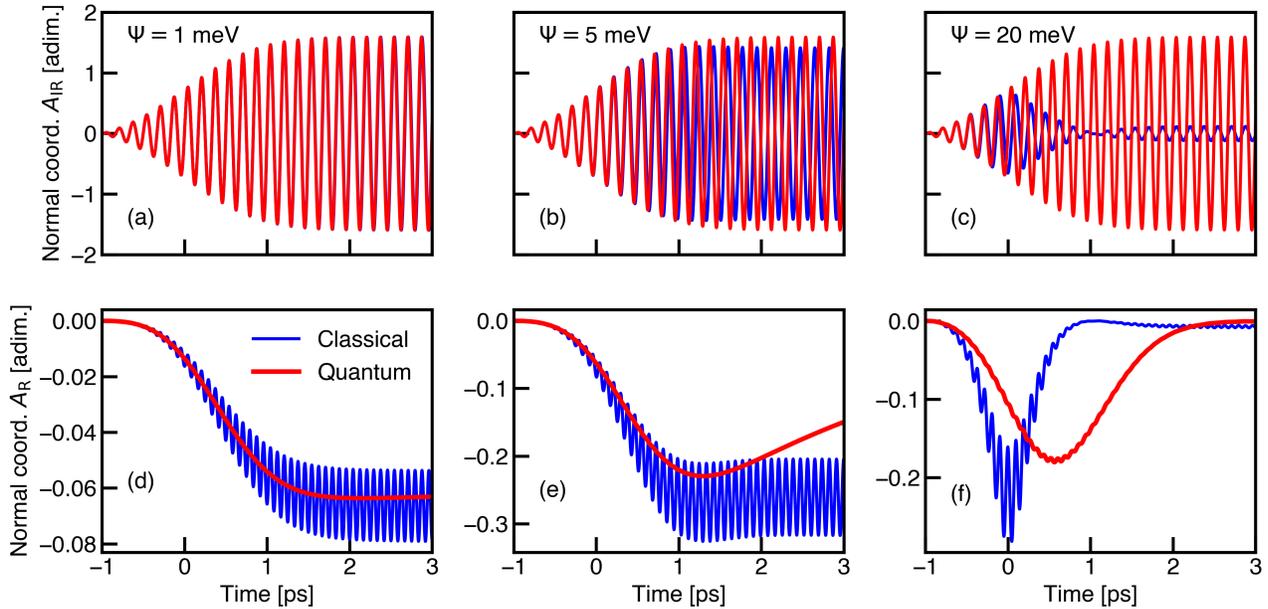}
\caption{\label{fig:QvsCl2}
Dynamics of the normal coordinate $Q_{\rm IR}$ of the IR-active mode for
third-order anharmonic coupling constant $\Psi = 1$, 5, and 20 meV [(a), (b),
and (c), respectively], as obtained from the time-propagation of the classical
EOM (blue, Eqs.~\eqref{eq:nlp2} and \eqref{eq:nlp2b}) and from the quantum EOM
(red, Eqs.~\eqref{eq:nlpQ} and \eqref{eq:nlpQb}) of the two-phonon model.
(d-f) Dynamics of the normal coordinate $Q_{\rm R}$ of the Raman-active mode
corresponding to the same anharmonic coupling $\Psi$ as in (a-c). The result of
the classical (quantum) EOM  is reported in blue (red). }
\end{center}
\end{figure*}

Having illustrated in Sec.~\ref{sec:ppclass} the classical approximation to the
lattice EOM, we proceed here to discuss the inclusion of quantum nuclear
effects on the dynamics. To address this task, we make use of the following
identity: 
\begin{align}\label{eq:AAq} \langle\hat Q_{\bq \nu }(t)\hat Q_{\bq' \nu' }(t)
\rangle = [2 n_{\bq\nu}(t) + 1] \delta_{-\bq\bq'}\delta_{\nu\nu'}\quad.
\end{align}
Here $n_{\bq\nu}(t)$ is the (incoherent) population of the phonon ${\bq\nu}$ at
time $t$ which can be directly obtained from the solution of the TDBE
(Sec.~\ref{sec:TDBE}).  Substitution of Eq.~\eqref{eq:AAq} into
Eq.~\eqref{eq:EOMpp} yields the quantum EOM for an anharmonic lattice in
presence of a THz field: 
\begin{align}\label{eq:EOMpp_Q} \D_t^2 Q_{\bq \nu} + &\omega_{\bq\nu}^2 Q_{\bq
\nu} = \delta_{\bq0} {\bf C}_\nu \cdot{\bf E}(t) \\&
- \frac{\omega_{\bq\nu}}{\hbar } \sum_{\bq'\nu'}  \Psi^{(3)}_{\substack{\nu
  \nu' \nu' \\ -\bq\bq'-\bq'}} [2 n_{\bq'\nu'}(t) +1] \nonumber \quad.
\end{align}
From the inspection of the phonon-phonon coupling matrix element $\Psi^{(3)}$,
the symmetry requirements for which the anharmonic corrections to
Eq.\eqref{eq:EOMpp_Q} become important can be easily deduced by elementary
group-theoretical considerations
\cite{cammarata_phononphonon_2019,Ravichandran2020}.  Namely, a prerequisite
for the matrix elements $\Psi^{(3)}$ to be finite is that the direct product of
the irreducible representations of the phonons involved in the scattering
process contains the totally-symmetric representation of the group $\Gamma_1$,
that is, $\Gamma_{\bq\nu} \otimes \Gamma_{\bq'\nu'} \otimes \Gamma_{-\bq'\nu'}
\supseteq \Gamma_1$.  In particular, because the phonons $\bq'\nu'$ and
$-\bq'\nu'$ belong to the same irreducible representation, the direct product
$\Gamma_{\bq'\nu'}\otimes\Gamma_{-\bq'\nu'}$ equals the totally symmetric
representation $\Gamma_1$. Therefore, the matrix elements $\Psi^{(3)}$ differs
from zero only if $\Gamma_{\bq\nu}$ contains (or is equal to) $\Gamma_1$.  This
symmetry requirement is violated for IR-active phonons or more generally for
phonons involving low-symmetry motion of the atoms.  The EOM for the normal
coordinates corresponding to these modes reduces to Eq.~\eqref{eq:EOMIR} and it
is unaffected by (third-order) lattice anharmonicities.  The symmetry condition
is satisfied by fully symmetric Raman-active modes ($A_{1g}$) for which the
anharmonic term in the Eq.~\eqref{eq:EOMpp_Q} remains finite. More generally,
other combinations of phonon symmetries allow for non-vanishing phonon-phonon
coupling matrix elements, and can therefore be driven by the mechanism in
Eq.~\eqref{eq:EOMpp_Q}.  In this case, the EOM resembles the problem of the
forced harmonic oscillator which --  for slow  variations for the phonon
occupation $n_{\bq'\nu'}(t)$ of the driven mode -- can be approximately solved
to yield an explicit expression for the time-dependent normal coordinate
$Q^{\rm R}_{\nu}$ of the Raman mode: 
\begin{align}\label{eq:R} Q^{\rm R}_{\nu}(t)  = - (\hbar \omega_{\nu})^{-1}
\sum_{\bq'\nu'}  \Psi^{(3)}_{\substack{\nu \nu' \nu'\\ 0\bq'-\bq'}} [2
n_{\bq'\nu'}(t) +1] \quad.  \end{align}
The expression for the nuclear displacement $\Delta {\boldsymbol \tau}_{\k p}$
is recovered by combining Eqs.~\eqref{eq:tau} and \eqref{eq:R}.  This mechanism
is the quantum counterpart of ionic Raman scattering and nonlinear phonon
rectification, typically described within the classical harmonic oscillator
model.  Several considerations can be immediately deduced from
Eq.~\eqref{eq:R}, which reveal the differences between a quantum and classical
description of the coherent lattice dynamics: (i) The displacement of the
lattice along totally symmetric modes (or along other modes complying with the
symmetry requirements outlined above) results from the enhancement of the
phonon population $n_{\bq\nu}$, rather than from the coherent phonon amplitude.
This indicates that ionic Raman scattering is an inherently {\it incoherent}
process. In other words, it results from the enhanced population of
strongly-coupled modes (as, e.g., driven IR-active modes), but it does not
necessarily requires the excitation of coherent phonons.  (ii) The displacement
of the lattice along totally-symmetric Raman modes contains a zero-point motion
(ZPM) component $Q^{\rm ZPM}_{\nu} = - (\hbar \omega_{\nu})^{-1}
\sum_{\bq'\nu'}  \Psi^{(3)}_{\substack{\nu \nu' \nu' \\ 0 \bq'\bq'}}$, which
should be present even at thermal equilibrium and zero temperature. $ Q^{\rm
ZPM}_{\nu}$ quantifies the displacement of the nuclear wave packets from
thermal equilibrium (i.e., from $\Delta {\boldsymbol \tau}_{\k p}=0$) due to
anharmonicity.  (iii) The time scales for light-induced structural
displacements due to ionic Raman scattering are limited by he characteristic
anharmonic lifetime of the driven IR-active mode.  In crystalline solids,
three-phonon scattering processes are the leading contribution to the mode
lifetime and they are accounted for by the collision integral $\Gamma^{\rm
pp}(t)$ within the framework of the Boltzmann equation
(Eq.~\eqref{eq:BTEexact}). 

\section{Two phonon model}\label{sec:model}
To illustrate the key differences between a classical and quantum formulation
of the coherent dynamics, we solved the classical [Eq.~\eqref{eq:EOMpp_Q}] and
quantum EOM [Eq.~\eqref{eq:nlp}]  for a model system consisting of two coupled
phonon modes.  In particular, we consider an IR-active mode with frequency
$\omega_{\rm IR} = 6$~THz, an electromagnetic field with frequency
$\omega=\omega_{\rm IR}$, and a fully-symmetric Raman-active mode with
frequency $\omega_{\rm R} = 0.8\omega_{\rm IR}$ coupled to the IR mode via
third-order anharmonicities.  Fourth-order anharmonicities are neglected in the
following.  Symmetry requires the coupling matrix elements to satisfy
$\Psi_{\rm IR,IR,R}^{(3)} = \Psi^{(3)}_{\rm IR,R,IR} =\Psi^{(3)}_{\rm R,IR,IR}
\equiv \Psi \neq 0$ and $\Psi_{\rm IR,R,R}^{(3)} =  \Psi^{(3)}_{\rm R,IR,R}
=\Psi^{(3)}_{\rm R,R,IR} = 0 $.  Correspondingly, the classical EOM
[Eq.~\eqref{eq:nlp}] for the normal coordinates of the IR- and Raman-active
modes, $Q_{\rm IR}$  and $Q_{\rm R}$, reduces to: 
\begin{align}\label{eq:nlp2} \D_t^2 Q_{\rm IR} + \omega_{\rm IR}^2 Q_{\rm IR}
&= { C}_{\rm IR} { E_0}(t) + g_{\rm IR}  Q_{\rm R } Q_{\rm IR }  \\ \D_t^2
Q_{\rm R} + \omega_{\rm R}^2 Q_{\rm R}    &= g_{\rm R}  Q_{\rm IR }^2 \quad,
\label{eq:nlp2b} \end{align}
where $C_{\rm IR}$ is defined as in Eq.~\eqref{eq:EOMIR}, and we introduced the
effective anharmonic coupling constants $g_{\rm IR} = - 2 \hbar^{-1}\omega_{\rm
IR} \Psi $ and $g_{\rm R} = - \hbar^{-1}\omega_{\rm R}  \Psi$.  More generally,
the effective coupling constant $g$ is related to the phonon-phonon coupling
matrix element via $g_{\nu\nu'\nu''} =  -\hbar^{-1} \omega_\nu
\Psi_{\nu\nu'\nu''}^{(3)} $.  From similar considerations, we deduce from
Eq.~\eqref{eq:EOMpp_Q} the quantum EOM for the two-phonon model: 
\begin{align}\label{eq:nlpQ} \D_t^2 Q_{\rm IR} + \omega_{\rm IR}^2 Q_{\rm IR}
&= { C}_{\rm IR} { E_0} f(t)  \\ \D_t^2 Q_{\rm R} + \omega_{\rm R}^2 Q_{\rm R}
&= 2 g_{\rm R}   n_{\rm IR} (t) \label{eq:nlpQb} \end{align}
where $n_{\rm IR}$ is the time-dependent phonon population of the IR-active
mode, which is obtained from the solution of the time-dependent Boltzmann
equation Eq.~\eqref{eq:BTEexact}.  Here, we omitted the zero-point motion
contribution  $ Q^{\rm ZPM}_{\nu}$ to the normal coordinate, as its effect is
limited to a renormalization of the equilibrium crystal structure.
Additionally, we note that the back-coupling of Raman mode $ Q_{\rm R}$ on the
IR mode $ Q_{\rm IR}$ is absent in the quantum approach, as a consequence of
the identity in Eq.~\eqref{eq:AAq}.  We solved numerically the classical EOM
[Eqs.~\eqref{eq:nlp2} and \eqref{eq:nlp2b}] by time-stepping the time
derivative via the Heun's finite-difference algorithm, whereas the quantum EOM,
Eqs.~\eqref{eq:nlpQ} and \eqref{eq:nlpQb}, have been solved simultaneously to
Eq.~\eqref{eq:BTEexact} to account for the increased population of IR phonons
due to IR absorption.  The field intensity and IR cross section are evaluated
considering realistic parameter values [see Table~\ref{tab:param}], and we
considered a pulse duration of $1$~ps. 

In Fig.~\ref{fig:QvsCl2}~(a-c), we compare the dynamics of the normal
coordinate $Q_{\rm IR}$ of the IR-active mode obtained from  the solution of
the classical (blue) and quantum (red) EOM, for increasing anharmonic coupling
constant $\Psi$.  For low anharmonic coupling (Fig.~\ref{fig:QvsCl2}~(a)), the
classical and quantum dynamics of the IR mode coincide, and they follow a trend
similar to the one illustrated in Fig.~\ref{fig:IR} for an ideal harmonic
lattice.  The dynamics of the IR mode as obtained from the quantum EOM is
unaffected by the value of $\Psi$, whereas the result of the  classical EOM
exhibits a dependence on the anharmonic coupling constant, which leads to an
increasing discrepancy with the quantum EOM in the strong coupling regime
(Figs.~\ref{fig:QvsCl2}~(b) and (c)). This effect can be attributed to the
influence that the Raman mode exerts on the IR-mode dynamics (second term in
the right-hand side of Eq.~\eqref{eq:nlp2}), which is suppressed in the quantum
EOM (right-hand side of Eq.~\eqref{eq:nlpQ}).  

The dynamics of the normal coordinate $Q_{\rm R}$ of the Raman mode is
illustrated in Figs.~\ref{fig:QvsCl2}~(d-f) for increasing anharmonic coupling
constant $\Psi$.  On short time scales, the normal coordinate  $Q_{\rm R}$
obtained from the quantum EOM exhibits a progressive shift away from the
initial value $Q_{\rm R}=0$, whereas on longer time scales, $Q_{\rm R}$ returns
to zero on timescales that are dictated by the phonon-phonon coupling strength.
This trend can be ascribed to the changes in phonon occupation  of the
IR-active mode, illustrated in Fig.~\ref{fig:TDBE_IR}, whose rise and decay
stems from IR absorption  and phonon-phonon scattering, respectively.  This
behaviour is the quantum counterpart of the ionic Raman scattering process
\cite{subedi_theory_2014,subedi_proposal_2015}.  While on the hand the results
of the classical (blue) and quantum EOM (red) follow qualitatively similar
trends, the classical results leads to periodic oscillations of the normal
coordinates.  This behaviour is further discussed in Sec.~\ref{sec:QvsC}.  The
agreement between quantum and classical picture can further be ameliorated
through the inclusion of a phenomenological damping term to account for the
decoherence of the coherent phonon amplitude.  Interestingly, the shift of the
normal coordinate $Q_{\rm R}$ [Eq.~\eqref{eq:R}] depends exclusively on (i) the
phonon-phonon coupling matrix elements $\Psi$ and (ii) the instantaneous
population of all phonons. These points suggest that it may be possible to
systematically assess and quantify light-induced structural distortions via
calculations of the third-order force constants of the IR activity alone, i.e.,
without the need for an explicit solution of the EOM of the lattice. 

\subsection{Classical vs quantum nuclear dynamics}\label{sec:QvsC}
To illustrate to which extent the coherent motion of the nuclei relates to the
incoherent dynamics of the lattice one can note that the phonon occupations
$n_\nu$ of zone-center phonons can be expressed as: 
\begin{align} \label{eq:cl2q} \langle \hat{Q}_{\nu} \hat{Q}_{\nu} \rangle = 2
n_{\nu} + 1 \quad.  \end{align}
Neglecting zero-point motion and quantum nuclear effects ($\langle
\hat{Q}_{\nu} \hat{Q}_{\nu} \rangle \simeq  Q_\nu ^2$), an approximate relation
between phonon number and normal coordinate can be deduced: $Q_\nu ^2 \simeq 2
n_\nu$.  In absence of decoherence or lattice anharmonicities, the amplitude of
THz-driven coherent nuclear motion can thus be approximately related to the
number of IR-active modes excited in the system. 

\begin{figure}[t] \begin{center}
\includegraphics[width=0.42\textwidth]{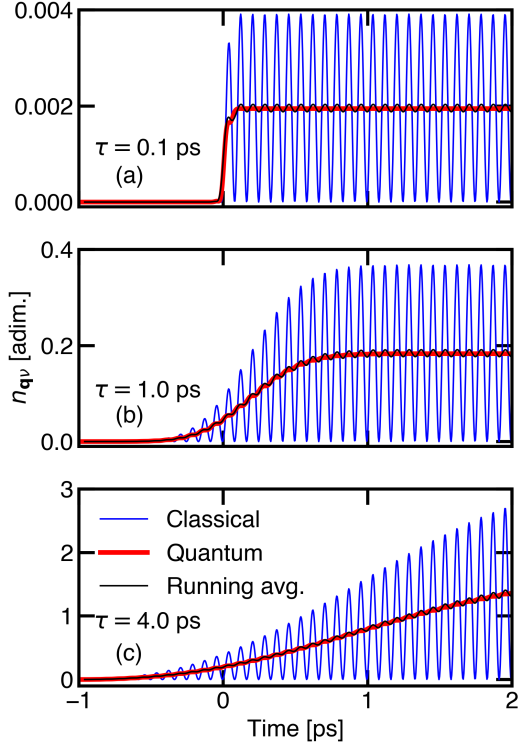} \caption{\label{fig:IRn}
Comparison between the phonon number $n_\nu$ (red), obtained from the solution
of the TDBE [Eq.~\eqref{eq:BTEexact}] and from the classical approximation
$n_\nu \simeq Q_\nu^2/2$ for  a pulse duration  $\tau = 0.1$ (a), 1 (b), and
10~ps (c).  For comparison, the running average of classical approximation is
included (black line) and it reveals excellent agreement with the quantum
results.  } \end{center} \end{figure}

We illustrate this point by  comparing in Fig.~\ref{fig:IRn} the phonon
occupation $n_\nu$ and corresponding classical approximation $Q_\nu ^2 /2$ for
an IR active phonon resonant to a THz field obtained from the numerical
solution of  Eqs.~\eqref{eq:EOMIR} and \eqref{eq:BTEexact}.  We consider a
frequency $\w_{\rm IR} = 6$~THz, and pulse durations $\tau = 0.1$, 1, and 10~ps
[(a), (b), and (c), respectively].  The monotonic increase of the phonon
occupations (red) reflects phonon emission processes resulting from the IR
absorption in absence of dissipation, decoherence, and anharmonicities.  The
classical approximation to the phonon number, conversely, exhibits an
oscillatory character, where the oscillation center coincides with the exact
(quantum) phonon number. In this case, the oscillations reflects the conversion
of kinetic and potentials energy (and vice versa).  This artifact can be
eliminated by considering the running average of the classical oscillation
amplitudes (black line in Fig.~\ref{fig:IRn}), which agrees well with the
phonon populations derived from the solution of the TDBE. 

\section{Conclusions}\label{sec:conc} 
We discussed the coherent structural dynamics of solids subject to the
interaction with strong fields and its theoretical foundation.  Our approach is
based on the solution of the Heisenberg equation for the nuclear displacement
operator, and it enables to derive a generalized EOM for coherent dynamics of
the crystalline lattice in presence of several interaction mechanisms.  The
lattice EOM can be solved exactly in some limiting case (e.g., for an harmonic
lattice interacting with a field), and it provides a way to systematically
include quantum nuclear effects in presence of anharmonicities. Additionally,
all known mechanisms for the excitation of coherent phonons can be accounted
for within a unified theoretical framework.  Classical approximations to the
coherent lattice dynamics (non-linear phononics model) can be immediately
deduced by neglecting the quantum character of the nuclei. 

These findings shed light on the origin of light-induced structural control in
THz-driven crystals.  In particular, the structural distortion of the lattice
along fully-symmetric Raman active modes  (ionic Raman scattering) emerges as
an {\it incoherent} process that results from the enhanced population of
anharmonically-coupled  IR-active modes.  The incoherent nature of the ionic
Raman scattering mechanism suggests that indirect control of the
non-equilibrium phonon population can provide a new route to engineer transient
structural changes in solids.  This picture corroborates and generalizes the
finding of earlier studies, whereby coherent structural dynamics was predicted
on the basis of the classical EOM.  Our findings further predict a zero-point
motion displacements of the nuclear wave packets along Raman-active modes which
is a fully quantum effect. This effect is not captured by classical
approximations, and it should occur even in absence of external perturbations.
Several of the analytical results deduced in this work are amenable for fully
ab-initio simulations. Overall, our findings establish a rigorous theoretical
foundation for  the light-driven coherent lattice dynamics and for the
widely-employed semi-classical models in the field of nonlinear phononics.  Our
approach is further suitable to account for novel phenomena that can trigger
the coherent dynamics of the lattice, such as, e.g., coherent phonon excitation
due  spin-phonon coupling in light-driven ferro-magnets. 

\section*{Acknowledgement(s)}
This project has been funded by the Deutsche Forschungsgemeinschaft (DFG) --
Projektnummer 443988403.  Discussions with Mariana Rossi, Davide Sangalli, and
Dominik Juraschek are gratefully acknowledged. 

\appendix

\section{Commutators relations for  $\hat{Q}_{\bq\nu}$ and $\hat{P}_{\bq\nu}$}\label{sec:commutators}
In this appendix, we summarize the commutation relations for
the operators $\hat{Q}_{\bq\nu} = \hat{a}_{\bq\nu} + \hat{a}^\d_{-\bq\nu}$ and 
$\hat{P}_{\bq\nu} = \hat{a}_{\bq\nu} - \hat{a}^\d_{-\bq\nu}$, which are  
required for the derivation of the  EOM of the lattice: 
\begin{align} 
[\hat{Q}_{\bq\nu}, \hat{Q}_{\bq'\nu' }] &= [\hat{P}_{\bq\nu}, \hat{P}_{\bq'\nu' }] = 0 \label{eq:cA1} \quad, \\
[\hat{P}_{\bq\nu}, \hat{Q}_{\bq'\nu' }] &= 2 \delta_{-\bq\bq'} \delta_{\nu\nu'}  \label{eq:cB1}\quad.  
\end{align}
Equations~\eqref{eq:cA1} and \eqref{eq:cB1} follow directly from the commutation 
relations for the bosonic operators $[ \hat{a}_{\bq\nu} ,  \hat{a}_{\bq'\nu'}^\d] =\delta_{\bq\bq'}\delta_{\nu\nu'}$ 
and  $[ \hat{a}_{\bq\nu} ,  \hat{a}_{\bq'\nu'} ] = [ \hat{a}_{\bq\nu}^\d ,  \hat{a}^\d_{\bq'\nu'} ] =0$. 
The commutators with the Hamiltonians specified in Eqs.~\eqref{eq:Hph}, \eqref{eq:Hpp},  \eqref{eq:Heph}, and \eqref{eq:HIR} 
can be easily derived by application of Eqs.~\eqref{eq:cA1} and \eqref{eq:cB1}: 
\begin{align}
[\hat{Q}_{\bq\nu}, \hat{H}_{\rm ph}] &= \hbar\w_{\bq\nu}\hat{P}_{\bq\nu} \label{eq:cAHph} \\
[\hat{P}_{\bq\nu}, \hat{H}_{\rm ph}] &= \hbar\w_{\bq\nu}\hat{Q}_{\bq\nu} \label{eq:cBHph} \\ 
 [\hat{Q}_{\bq\nu}, \hat{H}_{\rm IR}] &= 0 \\
 [\hat{Q}_{\bq\nu}, \hat{H}^{(3)}_{\rm pp}] &= 0 \\
 [\hat{Q}_{\bq\nu}, \hat{H}^{(4)}_{\rm pp}] &= 0 \\
 [\hat{Q}_{\bq\nu}, \hat{H}_{\rm eph}] &= 0 \\ 
 [\hat{P}_{\bq\nu}, \hat {H}_{\rm IR}  ]
&= - 2 e \delta_{\bq0}  \left( M_0 N_p \right)^{\frac{1}{2}} l_{ \bq  \nu } {\bf E}(t) \cdot {\bf F}_\nu 
\label{eq:commIR} \\ 
 [\hat{P}_{\bq\nu}, \hat{H}_{\rm pp}^{(3)}] &= \label{eq:commPP}
\sum_{\bq'\bq''}\sum_{\nu'\nu''} \Psi^{(3)}_{\substack{\nu\nu'\nu'' \\ -\bq\bq'\bq''}} \hat Q_{\bq' \nu' } \hat Q_{\bq'' \nu'' } \\
 [\hat{P}_{\bq\nu}, \hat{H}_{\rm pp}^{(4)}] &= \label{eq:commPP4}
\frac{1}{3}\sum_{\substack{{\bq'\bq''\bq'''}\\{\nu'\nu''\nu'''}}} 
\Psi^{(4)}_{\substack{\nu\nu'\nu''\nu''' \\ -\bq\bq'\bq''\bq'''}} \hat Q_{\bq' \nu' } \hat Q_{\bq'' \nu'' } \hat Q_{\bq''' \nu''' }  \\
 [\hat{P}_{\bq\nu}, \hat{H}_{\rm eph}] &= 
 N_p^{-\frac{1}{2}}
\sum_{mn\bk}  g_{mn}^\nu(\bk,-\bq) \hat c^\d_{m\bk-\bq} \hat c_{n\bk} \label{eq:commeph}
\end{align}
In Eq.~\eqref{eq:commIR}, the vector ${\bf F}_\nu=  \sum_{\k} { {{\bf Z}^{\text{*}}_\k } \cdot{\bf e}^\k_{\nu} } \, M_{\k}^{-\frac{1}{2}}$
is the IR cross section of the phonon $\nu$ with ${\bq=0}$, which
differs from zero only for IR-active phonons.
Equation~\eqref{eq:commPP} has been obtained by taking advantage of the 
symmetry properties of the third- and fourth-order 
phonon-phonon coupling matrix elements 
which are inherited by the corresponding force constant matrices \cite{Hellman2013}. 

\section{Effective Hamiltonian due to a THz field}\label{sec:HIR-deriv}

In this appendix, we report a derivation of the 
Hamiltonian term arising from the interaction with THz fields. 
In presence of an electromagnetic field, 
the kinetic energy of electrons and nuclei gets 
modified according to Peierl's substitution: 
\begin{align}
\hat {H}_{\rm IR} = \sum_{i} \frac{ e }{m_e } {\bf A} \cdot \hat{\bf p}_{i} 
- \sum_{\k p}  \frac{Z_\k e }{M_\k } {\bf A}\cdot \hat{\bf P}_{\k p} 
\end{align}
where $Z_\k$ is the charge of the $\k$-th nucleus, ${\bf A}$ is the vector potential, and $ \hat{\bf P}_{\k p}$ is the 
nuclear momentum. 
Making use of the relations: 
$\langle \Phi_f |  \hat{\bf P}_{\k p} | \Phi_i \rangle
 = i M_{\k } \omega  \langle\Phi_f |  \hat{\boldsymbol \tau}_{\k p} | \Phi_i \rangle $, and analogously 
for $\hat{\bf p}_{i}$, the effective Hamiltonian due to coupling with the field 
can be expressed as: 
\begin{align} \label{eq:HIRA0}
\hat{H}_{\rm IR} 
= e \sum_{i} {\bf E} \cdot \hat {\br}_i - e  \sum_{\k p} {Z_\k} {\bf E}\cdot \Delta \hat{\boldsymbol \tau}_{\k p} 
=  \hat{H}_{\rm IR}^{\rm (e)} +  \hat{H}_{\rm IR}^{\rm (n)} 
\end{align}
where ${\bf E} = -\D_t {\bf A}$.
Below we are interested in polar semiconductors with a band 
gap larger than the energy of incident phonons $\hbar\omega$. 
We can therefore restrict ourselves to consider 
matrix elements of the Hamiltonian $\hat{H}_{\rm IR}$ involving 
Born-Oppenheimer states $\ket{\Phi_i} = \ket{\chi_\nu}\ket{\Psi_s} $ 
and $\ket{\Phi_f} = \ket{\chi_\mu}\ket{\Psi_s} $, 
where $\ket{\chi_\nu}$ denote nuclear eigenstate. The 
electronic eigenstate $\ket{\Psi_s}$ remains unchanged and is 
assumed to be a Slater determinant. 
To derive an effective Hamiltonian acting only on the nuclear subspace, 
we project out the electronic degrees of freedom. 
We consider the expansion of the electronic eigenstate for small 
displacements up to second order:  
\begin{align} \label{eq:psiexp}
\ket{\Psi_{s}} 
&= \ket{\Psi^{(0)}_s} +\ket{\Psi^{(1)}_s} +\ket{\Psi^{(2)}_s}  \\
&= \ket{\Psi^{(0)}_s} + \sum_{\k p }\ket{\nabla_{\k p} \Psi_{s}} \cdot \Delta {\boldsymbol \tau}_{\k p} 
\end{align} 
where ${\nabla_{\k p} \Psi_{s}} = {\D { \Psi_{s}}}/{\D  \Delta {\boldsymbol \tau}_{\k p}} $ and 
similarly for ${\nabla_{\k p}  \nabla_{\k' p'}\Psi_{s}}$. 
Taking the expectation value of Eq.~\eqref{eq:HIRA0} with  Eq.~\eqref{eq:psiexp}, 
the zero-th order term in the expansion yields:
\begin{align} \label{eq:HIRA1}
\bra {\Psi^{(0)}_s} \hat{H}_{\rm IR}^{\rm (e)} \ket{\Psi^{(0)}_s} &=  
 e \sum_{i} {\bf E} \cdot \bra{\Psi^{(0)}_s}  \hat {\br}_i \ket{\Psi^{(0)}_s}
\\   
\bra {\Psi^{(0)}_s} \hat{H}_{\rm IR}^{\rm (n)} \ket{\Psi^{(0)}_s} 
&= -e  \sum_{\k p} {Z_\k} {\bf E}\cdot \Delta \hat{\boldsymbol \tau}_{\k p}. 
\end{align}
While the expectation value of the position operator $ \bra{\Psi^{(0)}_s}  \hat {\br}_i \ket{\Psi^{(0)}_s}$ 
is undefined, the electronic term vanishes for 
all cases  relevant for IR absorption because of 
the orthogonality of the nuclear eigenstates  $\bra{\chi_\nu} {\chi_\mu}\rangle = \delta_{\mu\nu}$. 
The first-order term yields: 
\begin{align} \label{eq:HIRA2}
 2 {\rm Re } & \bra {\Psi^{(0)}_s} \hat{H}^{\rm (e)}_{\rm IR} \ket{\Psi^{(1)}_s} 
= \\ \nonumber 
& e \sum_{i} \sum_{\k p} {\bf E} \cdot 
[ 2 {\rm Re} \bra {\Psi^{(0)}_s}   \hat {\br}_i  \ket{\nabla_{\k p} \Psi_{s}}]  \cdot  \Delta \hat{\boldsymbol \tau}_{\k p} \\
 2 {\rm Re } & \bra {\Psi^{(0)}_s} \hat{H}^{\rm (n)}_{\rm IR} \ket{\Psi^{(1)}_s}  = 0 \label{eq:HIRA2b}
\end{align}
Equation~\eqref{eq:HIRA2b} follows from $\bra {\Psi^{(0)}_s} {\Psi^{(1)}_s} \rangle = 0$. 
By considering an electronic eigenstate ${\Psi^{(0)}_s}$ in the form of a Slater determinant, 
the term in square bracket in Eq.~\eqref{eq:HIRA2} is recognized to be 
the electronic contribution to the Born effective change tensor: 
\begin{align}
{\bf Z}^{*{\rm (e)}}_{\k} =2 N_p^{-1} \sum_{n\bk} f_{n\bk}   {\rm Re} \bra{u_{n\bk}} \hat{\br} \ket{ \nabla_{\k p} u_{n\bk}}.
\end{align} 
where $f_{n\bk}$ is the electronic occupation of single-particle state ${n\bk}$, and ${u_{n\bk}}$ is the cell-periodic part 
of the Bloch eigenstate. 
By introducing the Born effective charge tensor 
${\bf Z}^{*}_{\k} = eZ_\k - {\bf Z}^{*{\rm (e)}}_{\k} $, the Hamiltonian $\hat{H}_{\rm IR}$ in 
Eq.~\eqref{eq:HIR} is recovered by combining Eqs.~\eqref{eq:HIRA1} and \eqref{eq:HIRA2}. 

\section{The collision integral due to IR absorption}\label{sec:appendix_gamma}
In this appendix, we briefly outline the derivation of 
the collision integral for IR absorption, reported in Eq.~\eqref{eq:GammaIR}.
The Heisenberg EOM 
for the bosonic operator $ \hat{a}_{\bq\nu}$ is:
\begin{align}
 {\D_t \hat{a}_{\bq\nu}(t)} &=- i\w_{\bq\nu} \hat{a}_{\bq\nu}(t)- {i}{\hbar^{-1}}
 [ \hat a_{\bq\nu}, \hat H_{\rm IR}](t) \label{eq:eomppA} 
\end{align}
where we consider a harmonic lattice interacting with a THz field, i.e., 
$\hat{H}= \hat{H}_{\rm ph} + \hat{H}_{\rm IR}$. A similar 
EOM holds for $ \hat{a}_{\bq\nu}^\d$.  Equation~\eqref{eq:eomppA} 
can be recast in the form of an integral equation as: 
\begin{align}\label{eq:aeomA}
 \hat{a}_{\bq\nu}(t) &=  e^{-i\w_{\bq\nu}t} \left[\hat{a}_{\bq\nu} (t_0) -  \frac{i}{\hbar}
\int_{t_0}^t d\tau [ \hat a_{\bq\nu}, \hat H_{\rm IR}](\tau) e^{i\w_{\bq\nu}\tau} \right]
\end{align} 
The phonon number operator at time $t$ is given by 
$ \hat{n}_{\bq\nu} (t) =  \hat{a}_{\bq\nu}^\d(t)  \hat{a}_{\bq\nu}(t) $, and it 
can be immediately derived from Eq.~\eqref{eq:aeomA} and its 
Hermitian conjugate:  
\begin{align}
  \hat{n}_{\bq\nu}(t) &=   \hat{n}_{\bq\nu} (t_0)
+ \frac{1}{4\hbar^2}
\left| \int_{t_0}^t d\tau [ \hat P_{\bq\nu}, \hat H_{\rm IR}](\tau) e^{-i\w_{\bq\nu}\tau} \right|^2
\end{align}
The collision integral $\Gamma_{\bq\nu}^{\rm IR}$ in Eq.~\eqref{eq:GammaIR} 
is promptly obtained from this expression by making use of the commutator in Eq.~\eqref{eq:commIR},  
taking the expectation value $\langle\cdots\rangle$, 
and  taking the derivative with  respect to time.

\end{document}